\documentclass{article}

\begin{document}

\leftline{\bf Boundary Effects in One-Loop Supergravity}
\vskip 0.3cm
\leftline{Giampiero Esposito$^{1,2}$ and 
Alexander Yu. Kamenshchik$^{3,4}$} 
\vskip 0.3cm
\noindent
{\it ${ }^{1}$INFN, Sezione di Napoli, Complesso Universitario di
Monte S. Angelo, Via Cintia, Edificio N', 80126 Napoli, Italy}
\vskip 0.3cm
\noindent
{\it ${ }^{2}$Dipartimento di Scienze Fisiche, Complesso Universitario
di Monte S. Angelo, Via Cintia, Edificio N', 80126 Napoli, Italy}
\vskip 0.3cm
\noindent
{\it ${ }^{3}$L.D. Landau Institute for Theoretical Physics, Russian
Academy of Sciences, Kosygina Str. 2, Moscow 117334, Russia}
\vskip 0.3cm
\noindent
{\it ${ }^{4}$Landau Network-Centro Volta, Villa Olmo, Via Cantoni 1,
22100 Como, Italy}
\vskip 0.3cm
Although Einstein's general relativity leads to a theory of
quantum gravity which is not perturbatively renormalizable,
the analysis of the semiclassical approximation remains
of crucial importance to test the internal consistency of any
theory of quantum gravity. For this purpose, it is necessary to
achieve a thorough understanding of the problem of boundary
conditions in the theory of quantized fields. 
Indeed, the path-integral representation of
the propagator, the general theory of the effective action,
and the recent attempts to define a quantum state of the
universe [1], provide three relevant examples where the
appropriate formulation of boundary conditions plays a crucial
role to obtain a well defined model of some properties of 
quantum gravity.

For gauge fields and gravitation, one may reduce the theory to 
its physical degrees of freedom by imposing a gauge condition
{\it before} quantization, or one may use the Faddeev--Popov
formalism for quantum amplitudes,  
or the extended-phase-space
Hamiltonian formalism of Batalin, Fradkin and Vilkovisky. 
Moreover, a powerful non-diagrammatic method to perform the
one-loop analysis is the one which relies on $\zeta$-function
regularization. This is a naturally occurring technique, since
semiclassical amplitudes involve definition and calculation of
determinants of elliptic, self-adjoint differential operators.
Once these choices are made, there are still many problems which
deserve a careful consideration. They are as follows.

(i) {\it Choice of background four-geometry}. This may be flat
Euclidean four-space, which is relevant for massless theories,
or the de Sitter four-sphere, which is relevant for
inflationary cosmology, or more general curved backgrounds.
The latter appear interesting for a better understanding of 
quantum field theory in curved space-time.

(ii) {\it Choice of boundary three-geometry}. This may consist
of two three-surfaces (e.g. two concentric three-spheres),
motivated by quantum field theory, or just one three-surface
(e.g. one three-sphere), motivated by quantum cosmology [1], 
or more complicated examples of boundary three-geometries.

(iii) {\it Choice of gauge-averaging functional}. For example,
one may study Lorenz or Coulomb gauge for Euclidean Maxwell
theory, or de Donder or axial gauge for gravitation,
or non-covariant gauges which
take explicitly into account extrinsic-curvature effects [2].

(iv) {\it Boundary conditions}. These may be Dirichlet, Neumann
or Robin for scalar fields; local or spectral for massless
spin-${1\over 2}$ fields; magnetic or electric for
Maxwell theory; local or spectral for spin-${3\over 2}$
potentials; completely invariant under infinitesimal 
diffeomorphisms on metric perturbations, or expressed in
terms of (complementary) projectors at the boundary, or
Robin on spatial perturbations and Dirichlet on normal
perturbations, or non-local, for linearized gravity.

Since recent calculations [2] have led to the
correct values of the one-loop divergences for spin-${1\over 2}$
fields, Euclidean Maxwell theory and Euclidean quantum gravity for
various choices of boundary conditions, including all ghost and
gauge modes whenever appropriate, 
the last open problem in this respect is the
evaluation of one-loop divergences in simple supergravity in the
presence of boundaries [2,3]. 

Simple supergravity is the simplest supersymmetric gauge theory
of gravitation. Its action functional consists of the
Einstein-Hilbert action, the massless gravitino action [1]
$$
I_{\psi,{\widetilde \psi}} \equiv {1\over 2}
\int_{M} \epsilon^{\mu \nu \rho \sigma} 
\left[{\widetilde \psi}_{\; \; \; \mu}^{A'} \;
e_{AA' \nu} \; D_{\rho} \; \psi_{\; \; \sigma}^{A}
-\psi_{\; \; \mu}^{A} \; e_{AA' \nu} \; D_{\rho} \;
{\widetilde \psi}_{\; \; \; \sigma}^{A'} \right] d^{4}x ,
\eqno (1)
$$
jointly with real and complex auxiliary fields (which
are necessary to close the supersymmetry algebra) and suitable 
boundary terms depending on the choices of boundary conditions 
at the bounding three-surfaces. With our notation, 
the gravitino potential is represented
by the pair of anticommuting, independent (i.e. not related
by any conjugation) spinor-valued one-forms 
$\psi_{\; \; \mu}^{A}$ and ${\widetilde \psi}_{\; \; \; \mu}^{A'}$.
These are obtained by contraction of spinor fields with the spinor
version of the tetrad as 
$$
\psi_{A \; \mu}=\Gamma_{\; \; \; AB}^{C'} 
e_{\; \; C' \; \mu}^{B} ,
\eqno (2)
$$
$$
{\widetilde \psi}_{A' \; \mu}=\gamma_{\; \; A'B'}^{C} 
e_{C \; \; \; \mu}^{\; \; B'} .
\eqno (3)
$$
The spinor fields $\Gamma$ and $\gamma$ occurring in Eqs. (2) and
(3) are the Rarita--Schwinger potentials. They are
subject to the {\it infinitesimal} gauge transformations 
$$
{\widehat \Gamma}_{\; \; \; BC}^{A'} \equiv 
\Gamma_{\; \; \; BC}^{A'}+\nabla_{\; \; \; B}^{A'} \nu_{C} ,
\eqno (4)
$$
$$
{\widehat \gamma}_{\; \; B'C'}^{A} \equiv 
\gamma_{\; \; B'C'}^{A}+\nabla_{\; \; B'}^{A} \mu_{C'} .
\eqno (5)
$$
For the spinor fields $\nu_{C}$ and $\mu_{C'}$ to be freely
specifiable inside the background four-manifold, the trace-free
part of the Ricci tensor has to vanish, jointly with the scalar
curvature. Hence the background is forced to be Ricci-flat.
 
Further restrictions are obtained on considering a {\it local}
description of $\Gamma$ and $\gamma$ in terms of a second 
potential. For example, on expressing locally the 
Rarita--Schwinger potentials as 
$$
\gamma_{\; \; A'B'}^{C}=\nabla_{BB'} 
\rho_{A'}^{\; \; \; CB} ,
\eqno (6)
$$
$$
\Gamma_{\; \; AB}^{C'} = \nabla_{BB'}  
\theta_{A}^{\; \; C'B'} ,
\eqno (7)
$$
one finds that the basic equations obeyed by the second potentials
$\rho$ and $\theta$ are invariant under infinitesimal gauge
transformations with gauge fields $\omega^{D}$ and $\sigma^{D'}$
if and only if 
$$
\psi_{AFLD} \omega^{D}=0 ,
\eqno (8)
$$
$$
{\widetilde \psi}_{A'F'L'D'} \sigma^{D'}=0 .
\eqno (9)
$$
With a standard notation, $\psi_{AFLD}$ and 
${\widetilde \psi}_{A'F'L'D'}$ are the anti-self-dual and
self-dual parts of the Weyl curvature spinor, respectively. 
Thus, to ensure unrestricted gauge freedom (except at the boundary)
for the second potentials $\rho$ and $\theta$, one is forced to
work in totally flat Euclidean backgrounds.
Such a restriction for massless gravitinos results from the form of the
action (1), which involves both $\psi_{\; \; \mu}^{A}$ and
${\widetilde \psi}_{\; \; \; \mu}^{A'}$, jointly with the local
description (6) and (7) and the form of any admissible set of 
boundary conditions.

Of course, the operator acting on Rarita--Schwinger
potentials is a Dirac operator. Within our framework,
which deals with positive-definite four-metrics, the Dirac operator
is a first-order elliptic operator which maps primed spinor fields
into unprimed spinor fields, and the other way around [4]. Thus,
the specification of the whole gravitino potential at the boundary
would lead to an over-determined problem. One thus has a choice of
spectral or local boundary conditions for Rarita--Schwinger
potentials, and they are both studied here in the presence
of three-sphere boundaries [3].

Spectral conditions reflect a choice which leads to a well posed
classical boundary-value problem. In other words, the massless
Rarita--Schwinger potential subject to gauge conditions and linearized
supersymmetry constraints is split into a regular part and a
singular part. The regular part is an infinite sum of modes 
multiplying harmonics having positive eigenvalues of the intrinsic
three-dimensional Dirac operator of 
the boundary [4]. By contrast,
the singular part is an infinite sum of modes multiplying harmonics
having negative eigenvalues of the intrinsic three-dimensional 
Dirac operator of the boundary. Such an identification relies
therefore on a {\it non-local} operation, i.e., the separation of the
spectrum of a first-order elliptic operator into its positive and
negative part [4].

When the corresponding semiclassical approximation of quantum theory
is studied, only half of the gravitino potential is set to
zero at the boundary. Bearing in mind the scheme described above,
one writes the spectral boundary
conditions on gravitino perturbations in the form
$$
\Bigr[\psi_{i(+)}^{A}\Bigr]_{\partial M}=0 ,
\eqno (10)
$$
$$
\Bigr[{\widetilde \psi}_{i(+)}^{A'}\Bigr]_{\partial M}=0 ,
\eqno (11)
$$
where the label $(+)$ denotes the part of the perturbation potential
corresponding to the regular part of the underlying classical theory.
To ensure invariance of the boundary conditions (10) and (11)
under the infinitesimal gauge transformations (4) and (5),
we require that
$$
\Bigr[\nabla_{\; \; \; B}^{A'} {\nu_{C}}_{(+)}\Bigr]_{\partial M}=0,
\eqno (12)
$$
$$
\Bigr[\nabla_{\; \; B'}^{A} {\mu_{C'}}_{(+)}\Bigr]_{\partial M}=0.
\eqno (13)
$$

As far as metric perturbations $h_{\mu \nu}$ are concerned, we
are interested in setting to zero at the boundary the spatial
perturbations, i.e. 
$$
\Bigr[h_{ij}\Bigr]_{\partial M}=0 .
\eqno (14)
$$
The six boundary conditions (14) are invariant under the
infinitesimal gauge transformations 
$$
{ }^{(\varphi)}h_{\mu \nu} \equiv h_{\mu \nu}
+\nabla_{(\mu} \; \varphi_{\nu)} ,
\eqno (15)
$$
if the whole ghost one-form $\varphi_{\nu}$ is set to zero
at the boundary:
$$
\Bigr[\varphi_{\nu}\Bigr]_{\partial M}=0 .
\eqno (16)
$$
At this stage, the boundary conditions on the normal components
$h_{00}$ and $h_{0i}$ are invariant under (15) if and only
if the whole gauge-averaging functional $\Phi_{\mu}$ is set to
zero at the boundary:
$$
\Bigr[\Phi_{\mu}(h)\Bigr]_{\partial M}=0 .
\eqno (17)
$$
A choice of gauge-averaging
functional which leads to self-adjoint operators on metric
perturbations is then the {\it axial}, i.e.
$$
\Phi_{\mu}(h) \equiv n^{\rho} h_{\mu \rho} ,
\eqno (18)
$$
where $n^{\rho}$ is the normal to the boundary. Since also the
proof of self-adjointness of squared Dirac operators
with spectral boundary conditions has been recently put on
solid ground [4], we have described so far a scheme where the boundary
conditions are completely invariant under infinitesimal gauge
transformations, the boundary operators on metric perturbations are
local and the differential operators on perturbative modes are
self-adjoint. Of course, the boundary conditions (14) and (17)
can be re-expressed in terms of tetrad vectors, and the one-loop
results coincide with those obtained from the metric 
formulation.

Another set of boundary conditions is instead motivated by work of
Luckock and Moss [2], where it is shown that the spatial tetrad 
$e_{\; \; \; \; \; i}^{AA'}$ and the projection
$\Bigr(\pm {\widetilde \psi}_{i}^{A'}-\sqrt{2} \;
{_{e}n_{A}^{\; \; A'}} \; \psi_{i}^{A}\Bigr)$ transform into
each other under half of the local supersymmetry transformations 
at the boundary. The resulting boundary conditions in one-loop
quantum cosmology about flat Euclidean four-space bounded by a
three-sphere with Euclidean normal ${_{e}n_{A}^{\; \; A'}}$
take the form 
$$
\sqrt{2} \; {_{e}n_{A}^{\; \; A'}} \; \psi_{i}^{A}
= \pm {\widetilde \psi}_{i}^{A'} \; {\mbox {at}} \;
\partial M ,
\eqno (19)
$$
jointly with the six boundary conditions (14) and the following
four boundary conditions on normal components of metric 
perturbations: 
$$
\left[{\partial h_{00} \over \partial \tau}
+{6\over \tau}h_{00}-{\partial \over \partial \tau}
\Bigr(g^{ij}h_{ij}\Bigr)\right]_{\partial M}=0 ,
\eqno (20)
$$
$$
\Bigr[h_{0i}\Bigr]_{\partial M}=0 ,
\eqno (21)
$$
where $g$ is the flat background four-metric, and $\tau$ is the
Euclidean-time variable, which plays the role of a radial coordinate.
Moreover, the ghost one-form for gravitons 
is subject to mixed boundary conditions, in that
the normal component $\varphi_{0}$ obeys Dirichlet conditions:
$$
\Bigr[\varphi_{0}\Bigr]_{\partial M}=0 ,
\eqno (22)
$$
and the tangential components obey Robin conditions: 
$$
\biggr[{\partial \varphi_{i}\over \partial \tau}
-{2\over \tau}\varphi_{i}\biggr]_{\partial M}=0 .
\eqno (23)
$$

The work of the authors [2,3] has found that, in the case of non-local 
boundary conditions (10) and (11) in the axial gauge, the contributions
of ghost and gauge modes vanish separately, and hence contributions to
the one-loop wave function of the universe reduce to those $\zeta(0)$ 
values resulting from transverse-traceless perturbations only. By contrast, when
the mixed boundary conditions (19)--(23) motivated by local supersymmetry
are imposed, the contributions of gauge and ghost modes do not cancel
each other. Both sets of boundary conditions lead to a non-vanishing
$\zeta(0)$ value, and spectral boundary conditions have also been studied
when two concentric three-sphere boundaries occur [2,3]. These results 
seem to point out that simple supergravity is not even one-loop finite 
in the presence of boundaries, apart from the case of those particular
backgrounds bounded by two surfaces where pure gravity alone is one-loop
finite. More recently, work by Deser and Seminara [5], devoted to
four-particle bosonic scattering amplitudes in $D=11$ supergravity, has
shown that no symmetries protect this ultimate supergravity from the
non-renormalizability of its lower-dimensional counterparts. This has been
proved by finding a non-vanishing two-loop counterterm.
It therefore seems that the issue of finiteness of quantum supergravity
is far from being settled, both in the presence [1--3] and in the absence [5]
of boundaries. Hopefully, the methods of modern spectral geometry [4]
will make it possible, in the years to come, to perform a more careful
assessment of all possible counterterms at various orders in perturbation
theory. Such an effort should yield an {\it a priori} understanding of
why the original expectations are, or are not, vindicated.
\vskip 0.3cm
\leftline{BIBLIOGRAPHY}
\vskip 0.3cm
[1] P.D. D'Eath, {\it Supersymmetric Quantum Cosmology}, 
Cambridge University Press 1996;

[2] G. Esposito, A.Yu. Kamenshchik and G. Pollifrone,
{\it Euclidean Quantum Gravity on Manifolds with Boundary}, Kluwer 1997;

[3] G. Esposito and A.Yu. Kamenshchik, PR D54 (1996) 3869;

[4] G. Esposito, {\it Dirac Operators and Spectral Geometry}, 
Cambridge University Press 1998;

[5] S. Deser and D. Seminara, PR D62 (2000) 080410.

\end{document}